\documentclass[twocolumn,preprintnumbers,superscriptaddress,nofootinbib,aps,prd,floatfix]{revtex4}

\usepackage{epsf,epsfig,subfigure,graphicx,amsmath,amssymb,cancel,slashed}
\usepackage{subfigure}
\usepackage{soul}
\usepackage[colorlinks=true,urlcolor=blue,anchorcolor=red,citecolor=darkgreen,filecolor=red,linkcolor=red,menucolor=blue]{hyperref}
\usepackage[usenames,dvipsnames]{xcolor}

\usepackage{afterpage}
\usepackage[utf8]{inputenc}

\allowdisplaybreaks

\newcommand{\newc}{\newcommand}
\newc{\gsim}{\lower.7ex\hbox{$\;\stackrel{\textstyle>}{\sim}\;$}}
\newc{\lsim}{\lower.7ex\hbox{$\;\stackrel{\textstyle<}{\sim}\;$}}
\newc{\gev}{\,{\rm GeV}}
\newc{\mev}{\,{\rm MeV}}
\newc{\ev}{\,{\rm eV}}
\newc{\kev}{\,{\rm keV}}
\newc{\tev}{\,{\rm TeV}}
\newc{\MHT}{$H_T^{\text{miss}}$}
\newc{\MET}{$\slashed{E}_T$}
\newc{\MTT}{$M_{T2}$}

\newc{\mz}{M_Z}
\newc{\mpl}{M_*}
\newc{\mw}{m_{\rm weak}}
\newc{\nr}[1]{N^c_R{}_{#1}}

\newcommand{\ubl}{\mathrm{U(1)}_{\mathrm{B-L}}}

\def\beq{\begin{equation}}
\def\eeq{\end{equation}}
\newcommand{\bea}{\begin{eqnarray}\begin{aligned}}
\newcommand{\eea}{\end{aligned}\end{eqnarray}}
\def\bitem{\begin{itemize}}
\def\eitem{\end{itemize}}

\definecolor{darkgreen}{rgb}{0,0.5,0}
\definecolor{goodyellow}{rgb}{0.9,0.7,0}

\begin{document}

\vspace{-1cm}

MIT-CTP/5412

\title{Glueballs in a Thermal Squeezeout Model}

\vskip 1.0cm
\author{Pouya Asadi}
\thanks{{\scriptsize Email}: \href{mailto:pasadi@mit.edu}{pasadi@mit.edu}; 
}
\affiliation{Center for Theoretical Physics, Massachusetts Institute of Technology, \\ Cambridge, MA 02139, USA.}
\author{Eric David Kramer}
\thanks{{\scriptsize Email}: \href{mailto:ericdavi@g.jct.ac.il}{ericdavi@g.jct.ac.il}; 
}
\affiliation{Jerusalem College of Technology, Jerusalem, Israel.}
\affiliation{Racah Institute of Physics, Hebrew University of Jerusalem, Jerusalem 91904, Israel.}
\author{Eric Kuflik}
\thanks{{\scriptsize Email}: \href{mailto:eric.kuflik@mail.huji.ac.il}{eric.kuflik@mail.huji.ac.il};
}
\affiliation{Racah Institute of Physics, Hebrew University of Jerusalem, Jerusalem 91904, Israel.}
\author{Tracy R. Slatyer}
\thanks{{\scriptsize Email}: \href{mailto:tslatyer@mit.edu}{tslatyer@mit.edu}; 
}
\affiliation{Center for Theoretical Physics, Massachusetts Institute of Technology, \\ Cambridge, MA 02139, USA.}
\author{Juri Smirnov}
\thanks{{\scriptsize Email}: \href{mailto:juri.smirnov@fysik.su.se}{juri.smirnov@fysik.su.se};
{\scriptsize ORCID}: \href{http://orcid.org/0000-0002-3082-0929}{0000-0002-3082-0929}}
\affiliation{Stockholm University and The Oskar Klein Centre for Cosmoparticle Physics, \\ Alba Nova, 10691 Stockholm, Sweden.}

\begin{abstract}
It has been shown that a first order confinement phase transition can drastically change the relic dark matter abundance in confining dark sectors with only heavy dark quarks. 
We study the phenomenology of one such model with a $Z'$ portal to Standard Model. We find that dark glueballs are long-lived in this setup and dilute the dark matter abundance after their decay to Standard Model. 
With this effect, the correct relic abundance is obtained with dark matter masses up to $\mathcal{O}(10^6)$~TeV. We find that while a part of the parameter space is already ruled out by direct detection and collider searches, there is still a broad space of viable scenarios that can be probed by future experiments.

\end{abstract}

\maketitle

\vskip 1cm

\newpage

\section{Introduction}
\label{sec:intro}

The particle nature of dark matter (DM) and the interactions controlling its dynamics are two of the major unanswered questions in the current standard model (SM) of particle physics. While the original idea of Weakly Interacting Massive Particles (WIMPs) \cite{Zeldovich:1965gev,Lee:1977ua,Steigman:1984ac,Steigman:2012nb} remains a viable possibility to date \cite{Leane:2018kjk}, many other intriguing dynamics have been studied as well.

If the ongoing experiments fail to find any DM particles in the conventional mass range of  thermal perturbative DM models with only DM-DM annihilations, i.e. $\mathcal{O}(100)$~TeV or lower, it is only natural to consider next-to-minimal dynamics that can open up new parameter space for higher masses (for heavier DM and unitary bounds from other perturbative $2 \to 2$ freezeout processes see Refs.~\cite{Kim:2019udq,Kramer:2020sbb}) . 
An interesting class of next-to-minimal DM models are confining dark sector scenarios, see for instance Refs.~\cite{Gudnason:2006yj,Alves:2009nf,Kilic:2009mi,Hambye:2009fg,Kribs:2009fy,Alves:2010dd,Bai:2010qg,Feng:2011ik,Fok:2011yc,Lewis:2011zb,Frigerio:2012uc,Buckley:2012ky,Appelquist:2013ms,Bhattacharya:2013kma,Cline:2013zca,Boddy:2014yra,Boddy:2014qxa,Hochberg:2014kqa,Appelquist:2015yfa,Antipin:2015xia,Soni:2016gzf,Harigaya:2016nlg,Mitridate:2017oky,Forestell:2017wov,DeLuca:2018mzn,Contino:2018crt,Gross:2018zha,Geller:2018biy,Beylin:2019gtw,Dondi:2019olm,Buttazzo:2019mvl,Landini:2020daq,Brower:2020mab,Baldes:2020kam,Contino:2020god,Jo:2020ggs,Beylin:2020bsz,Redi:2020ffc,Garani:2021zrr,Contino:2020tix,Asadi:2021yml,Asadi:2021pwo,Gross:2021qgx,Arakawa:2021wgz,Reichert:2021cvs,Cheng:2021kjg,Baldes:2021aph,Bensalem:2021qtj}, and Refs.~\cite{Kribs:2016cew,Cline:2021itd} for recent reviews. 

Using a novel approach, Refs.~\cite{Asadi:2021yml,Asadi:2021pwo} showed that the dark confinement phase transition (PT) of such models can substantially affect the DM abundance calculation, as we summarize in the next section. As a result of this ``thermal squeezeout" effect, the dark matter mass consistent with the relic density increases to masses far beyond the perturbative unitarity bound on thermal DM mass~\cite{Griest:1989wd,vonHarling:2014kha,Smirnov:2019ngs}.

The thermal squeezeout effect is independent of the details of the portal between SM and the dark sector. Nonetheless, a portal is still required between the two sectors to allow the dark glueballs to decay to SM particles. If these glueballs were stable, they would have become the dominant fraction of the DM in such setups~\cite{Redi:2020ffc}, and in the mass range where thermal squeezeout becomes relevant would lead to an overproduction of the DM abundance.

In this work we consider a specific model for the thermal squeezeout scenario found in Refs.~\cite{Asadi:2021yml,Asadi:2021pwo}, that includes a single flavor of heavy dark quarks charged under a confining SU(3) gauge group. 
To enable a decay of all dark glueballs to SM particles, we must either have particles charged both under SM and dark gauge groups (e.g. see Ref.~\cite{Mitridate:2017oky}), or a vector portal between the two sectors~\cite{Juknevich:2009gg}. Owing to its simplicity, we will focus on the latter and use a $\ubl$ portal in this work. 

We find that, for the ranges of dark quark masses and dark confinement scales we are studying, it is natural for the dark glueballs to be long-lived enough to eventually dominate the energy density of the universe after the PT, giving rise to an early matter-dominated epoch. 
Once the glueballs decay to SM particles, they can inject substantial entropy into the SM thermal bath that can further suppress the DM relic abundance. This, in turn, pushes the relic abundance line to even higher masses than predicted in Refs.~\cite{Asadi:2021yml,Asadi:2021pwo}.

\begin{figure}
    \centering
    \resizebox{0.8\columnwidth}{!}{
    \includegraphics{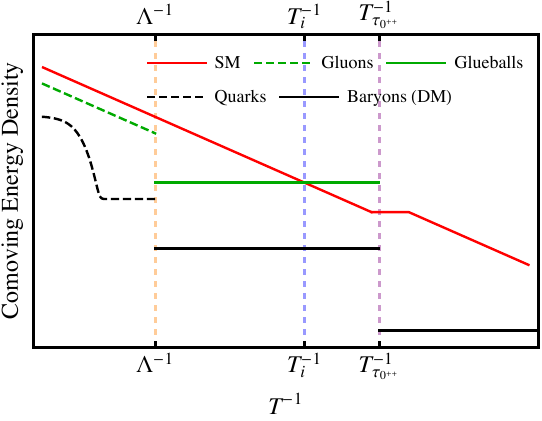}
    }
    \caption{ The schematic evolution of different relics abundances in our setup. We show the schematic contours for SM (red - neglecting changes in number of relativistic degrees of freedom), dark quarks and dark baryons (dashed and solid black lines), and dark gluons and glueballs (dashed and solid green lines). The PT epoch is marked by the dashed orange line. We have assumed the glueballs are long-lived enough that they take over the energy budget of the universe at some point (marked by the dashed blue line) and eventually decay to the SM at a lower temperature (dashed purple line). As a result, they inject significant amount of entropy into the SM bath that further suppresses the dark baryons abundance. Due to this injection, SM comoving abundance dilutes more slowly during this epoch. }
    \label{fig:evolution}
\end{figure}

A schematic evolution of different relics' abundances in our setup is shown in Fig.~\ref{fig:evolution}. 
The abundance of dark quarks is depleted during the PT due to the thermal squeezeout effect; the glueballs' abundance will also be suppressed during the confinement PT due to complicated dynamics of the PT as we will explain in this work. The glueballs then freeze out and, if they are long-lived enough, eventually take over the energy budget of the universe at $T=T_i$. They will eventually decay to SM particles at $T(t)=T(\tau_{0^{++}})\equiv T_{\tau_{0^{++}}}$ where $\tau_{0^{++}}$ is the lifetime of the lightest glueball, injecting substantial amount of entropy into the thermal bath that dilutes the DM abundance further. 
Our entropy injection setup can be thought of as a special case of WIMPs Without Weakness \cite{Asadi:2021bxp}.

This entropy injection has the most extreme effect when the glueballs decay roughly around the onset of BBN. 
We will focus on this scenario. 
We will also study other signals of this specific setup in ongoing experiments and find that despite large DM masses, there can still be potentially observable signals in a range of experiments.

In Sec.~\ref{sec:squeezeout} we introduce the model and discuss the squeezeout effect and entropy injection that dilute DM abundance. Phenomenology of the model is studied in Sec.~\ref{sec:pheno}. We conclude in Sec.~\ref{sec:conclusion}. 
In App.~\ref{sec:appxTherm} we use simple thermodynamic relations to provide an analytical approximation for the PT timescale. 
A dynamical approximation for the glueball abundance during the PT is presented in App.~\ref{sec:appxDyn}. Finally, in App.~\ref{sec:appxGB} we justify why keeping only the lightest glueball state in our analysis is a good approximation.


\section{The Model}
\label{sec:squeezeout}

We extend the Lagrangian of the SM with a confining SU(3), new dark quarks $q$, and a gauged $\ubl$
\begin{eqnarray}
\label{eq:L}
\mathcal{L} & \supset & \bar{q} (i\slashed{D} -m_q) q - \frac{1}{4} Z^{'\mu\nu}Z^{'}_{\mu\nu} - \frac{1}{4} G^{\mu\nu}G^{}_{\mu\nu},
\end{eqnarray}
where $Z^{'\mu\nu}$ ($G^{\mu\nu}$) is the field strength of the new $\ubl$ gauge group (dark SU(3)).\footnote{We assume right-handed neutrinos exist to ensure anomaly cancellation; they do not affect the rest of our study in this work.} The covariant derivative of the SM fields now includes the interaction with the new $\ubl$ gauge boson $Z'$. The mass and gauge coupling of this new gauge field are denoted by $m_{Z'}$ and $g_{Z'}$, respectively. 
We focus on the parameter space with $m_{Z'} \leqslant \Lambda$, where $\Lambda$ is the dark sector's confinement scale, to simplify the glueballs' decay calculation. 
It is shown that a vector boson portal like the $Z'$ in our model enables the decay of all glueballs \cite{Juknevich:2009ji,Juknevich:2009gg}. 
Another intriguing scenario arises in the case that the $Z’$ mass becomes small, which leads to long range interactions~\cite{diyann}.

With a $\ubl$ portal the couplings of all SM particles to the $Z'$ are determined. 
The dark quarks are vector-like fermions; thus, the anomaly constraints are guaranteed to be satisfied. We explicitly set the charges of dark quarks under $\ubl$ to $q_q=1$. The remaining degrees of freedom are (1) the heavy dark quark mass $m_q$, (2) the dark confinement scale $\Lambda$, (3) the $Z'$ mass $m_{Z'}$, and (4) its gauge coupling $g_{Z'}$. Our goal is to explore the space of these parameters and find the parts that explain the observed DM abundance. 

\begin{figure}
    \centering
    \resizebox{\columnwidth}{!}{
    \includegraphics{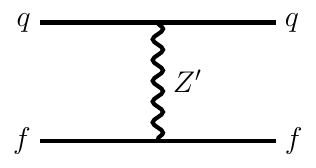}
    \includegraphics{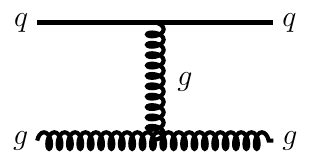}
    }\\
    \resizebox{\columnwidth}{!}{
    \includegraphics{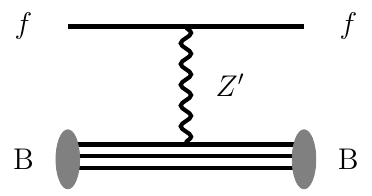}
    \includegraphics{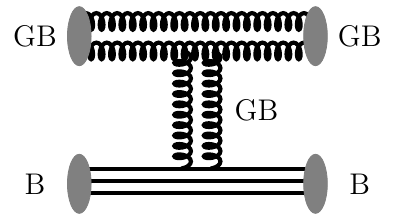}
    }
    \caption{Diagrams giving rise to heat exchange between different particles in the spectrum before (top) and after (bottom) the confinement PT. One can easily check that the top diagrams keep the dark quarks $q$ in equilibrium with SM fermions $f$ (top left) and with dark gluons $g$ (top right). This suggests the SM and dark gluons remain in kinetic equilibrium as well.
    After the PT the abundance of all the relics in the dark sector get substantially suppressed and heat exchange rates become much lower. 
    The most efficient diagrams for heat exchange after the PT are shown between the dark baryons and SM fermions (bottom left), and dark glueballs and baryons (bottom right). These diagrams can not maintain the kinetic equilibrium between SM and dark glueballs after the PT, see text for further details. 
    }
    \label{fig:keq}
\end{figure}

Fig.~\ref{fig:keq} (top row), shows the processes that keep the dark quarks in kinetic equilibrium with the SM and the dark gluons before the PT. 
The dark quarks remain in equilibrium with SM (via the top-left diagram in Fig.~\ref{fig:keq}) when the heat exchange rate is larger than the Hubble rate:
\begin{eqnarray}
    \label{eq:quarkSMkeq}
    \frac{\Gamma_h}{H(T)} \simeq \frac{n_f \langle \sigma v \rangle T/m_q }{g_*^{1/2} T^2/M_{\mathrm{pl}}} \gtrsim 1 ,
\end{eqnarray}
where $\Gamma_h$ is the heat transfer rate between dark quarks and SM fermion $f$, $n_f$ is SM fermion number density, $\langle \sigma v \rangle$ is the relevant cross section, the $T/m_q$ factor is inverse of number of interactions required for enough heat transfer \cite{Rubakov:2017xzr}, $g_*$ is the relativistic degrees of freedom at temperature $T$, and $M_{\mathrm{pl}}=1.2 \times 10^{19}$ is the Planck mass. Assuming $m_{Z'}$ is negligible compared to transferred energy in the top-left diagram of Fig.~\ref{fig:keq}, we find that at $T=\Lambda$
\begin{eqnarray}
    \label{eq:quarkSMkeqnumbers}
    \frac{\Gamma_h}{H(\Lambda)} &\approx& 10 \times \alpha^2_{Z'} \frac{m_q}{\Lambda} \frac{M_{\mathrm{pl}}}{\Lambda} \\
    &\approx & 10^{16} \times \alpha_{Z'}^2 \times \left(  \frac{m_q/\Lambda}{10^{2}} \right) \times \left(  \frac{\Lambda}{10^3~\mathrm{TeV}} \right)^{-1}. \nonumber
\end{eqnarray}
For the range of parameters studied in this work (see below), this equation suggests that for $\alpha_{\rm Z'} \gtrsim 10^{-8}$ the kinetic equilibrium will be maintained at the time of confinement PT. Similarly, we can check that, since dark gluons are massless, with an elastic scattering rate factor that scales as $\langle \sigma v \rangle \sim T^{-2}$, they maintain their equilibrium with dark quarks (and subsequently with SM particles) until the PT as well.
The gluon abundance tracks the equilibrium value while the quarks eventually freeze out and their freeze-out abundance can be calculated using usual perturbative techniques \cite{Asadi:2021pwo}. The process controlling their abundance is $qq \rightarrow gg$ and the relevant cross section was summarized in the appendix of Ref.~\cite{Asadi:2021pwo}.\footnote{The annihilation to SM via a $Z'$ is also allowed but we checked that it is sub-dominant to the aforementioned process.}

We consider heavy enough dark quarks such that (1) they freeze out before the confinement PT, and (2) the confinement transition is of first order \cite{Svetitsky:1982gs, Kaczmarek:1999mm,PhysRevD.60.034504,Aoki:2006we,Saito:2011fs}. In Refs.~\cite{Asadi:2021yml,Asadi:2021pwo} it was argued that the PT has a profound effect on the dark matter abundance. We start by summarizing this effect.

\subsection{Thermal squeezeout of dark matter - a summary}
\label{subsec:squeezeout}

At the onset of the dark confinement PT, bubbles of the confined phase nucleate inside the ambient sea of the deconfined phase. For these bubbles to nucleate efficiently slight supercooling is required. After their nucleation, bubbles grow since they are energetically favored. During this growth, latent heat is released, and the vicinity of the bubble heats back to $T \approx \Lambda$. It has been argued \cite{Witten:1984rs,Asadi:2021pwo} that the bubble growth rate is limited by the heat diffusion rate, which in turn is proportional to the amount of supercooling during the PT. For the small supercoolings that happen in our confinement PT the bubble walls reach a non-relativistic terminal velocity. Further numerical studies are required to improve our understanding of the bubble wall velocity.

\begin{figure}
    \centering
    \resizebox{\columnwidth}{!}{
    \includegraphics{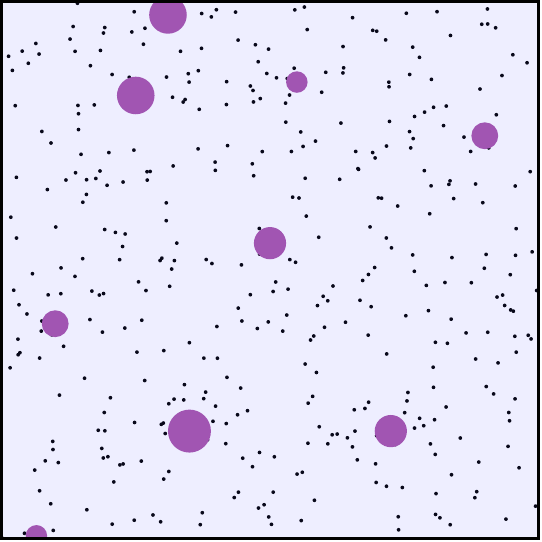}
    \includegraphics{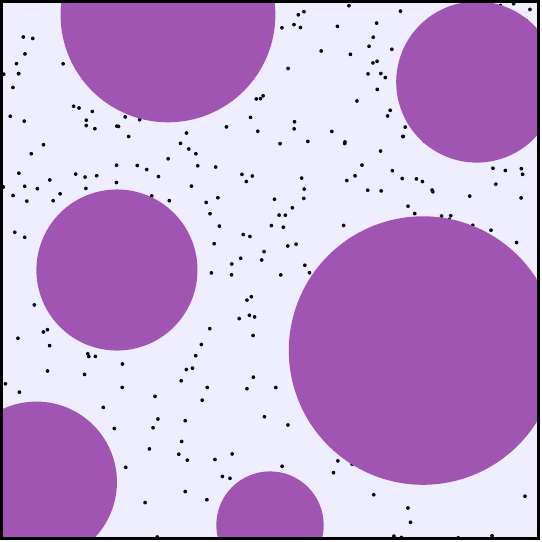}
    \includegraphics{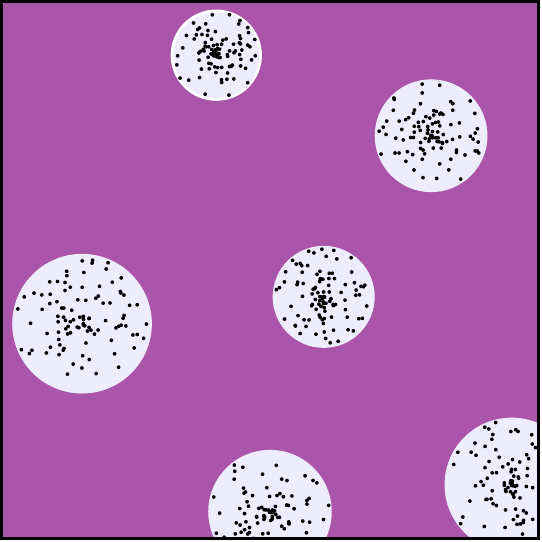}
    }\\
    \resizebox{\columnwidth}{!}{
    \includegraphics{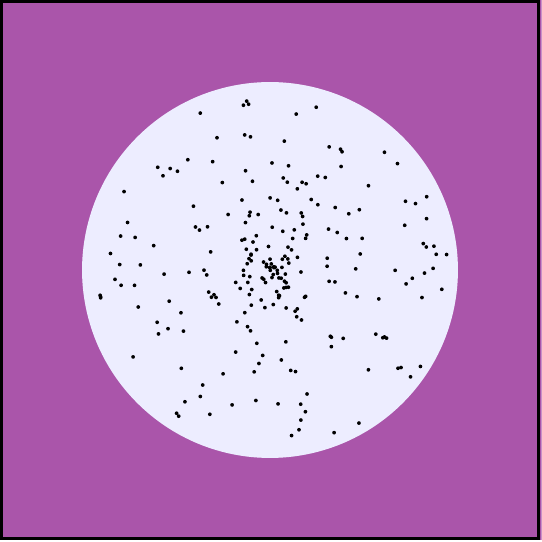}
    \includegraphics{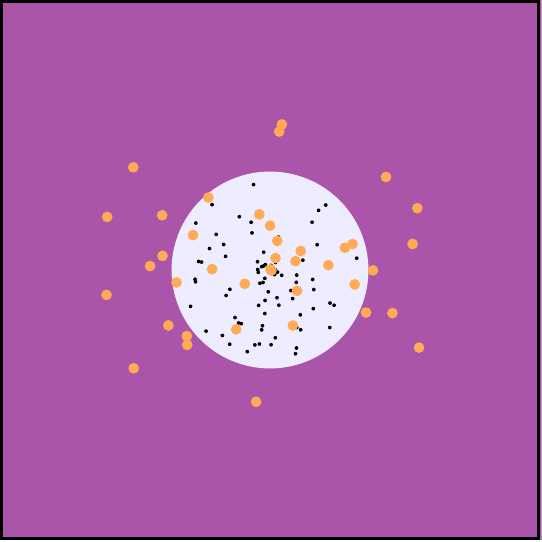}
    \includegraphics{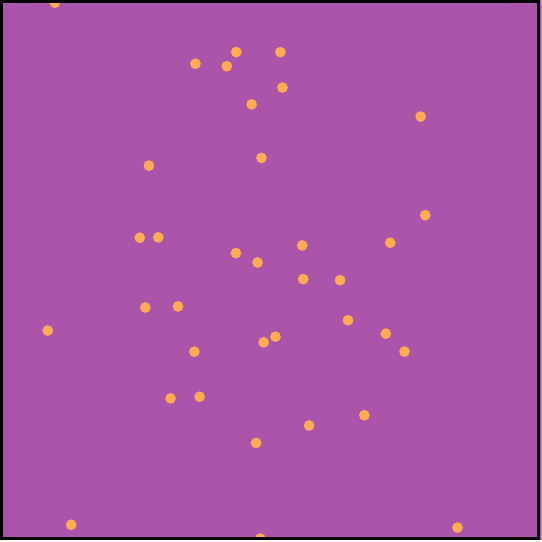}
    }
    \caption{Different stages of the confinement PT in the dark sector and its effect on the dark quarks. Figure is taken from Refs.~\cite{Asadi:2021yml,Asadi:2021pwo}. The PT starts with bubbles of the confined phase (dark purple) nucleating inside the ambient sea of the deconfined phase (light purple). As the bubbles grow, they push aside isolated quarks (black dots) and eventually sweep them inside isolated pockets of the deconfined phase after the percolation. As each pocket contracts, it brings the trapped quarks into interaction again. The quarks annihilate against each other into gluons/glueballs very efficiently and only a small fraction of them survive in the form of stable baryons (orange dots). }
    \label{fig:PT}
\end{figure}

The evolution of the PT and its effect on quarks are summarized in Fig.~\ref{fig:PT}. As the bubbles grow, they run into heavy dark quarks. An isolated quark sees an effectively infinite energy barrier as it encounters a bubble of the confined phase, so it gets pushed back into the deconfined phase and remains trapped there.

After bubbles percolate, the deconfined phase survives only in the form of isolated pockets surrounded by a sea of confined phase. In Ref.~\cite{Asadi:2021pwo}, we calculate the number density of quarks in these pockets and argue that they are still very diffused after the percolation. As these pockets contract, the quarks are brought back together and go through a second stage of annihilation that depletes their abundance.
We find that only a small fraction of these quarks form stable baryons during this squeezeout and move into the confined phase (the baryons are color-neutral and do not feel an infinite potential barrier when encountering the pocket wall). 
The surviving baryons eventually account for the DM observed today.

In Refs.~\cite{Asadi:2021yml,Asadi:2021pwo} it was also pointed out that the accidental net baryon number in the pockets can give rise to a lower bound on the survival rate from the pockets. 
In this work we assume this bound on the survival factor is saturated; 
this allows us to find the highest viable DM masses in this setup.

It is worth pointing out that the calculation of Ref.~\cite{Asadi:2021pwo} is merely the initiation of a closer look at the effect of a confinement PT on the DM abundance and more careful studies are in order to obtain more precise predictions for the DM abundance. In particular, the inferred relation between the DM mass and confinement scale in Ref.~\cite{Asadi:2021pwo} has a roughly one-order-of-magnitude error bar due to a number of systematic uncertainties. In this work we use the central value of this relation as a benchmark.

\subsection{Glueballs in the early universe}
\label{subsec:GBs}

In SU(3) Yang-Mills, there are twelve different glueballs (with different $J^{PC}$ numbers) that are stable in the absence of other interactions~\cite{Juknevich:2009gg}. 
The lightest of these glueballs, $J^{PC}=0^{++}$, has a mass of $m_{0^{++}} \approx 7 \Lambda$ (see Ref.~\cite{Mathieu:2008me} and the references therein). The mass of these states spans a range of $\sim 7-20~\Lambda$. 
In our analysis we only keep the lightest glueball ($0^{++}$); in App.~\ref{sec:appxGB} we argue that this is a good approximation. 

In Fig.~\ref{fig:keq} (bottom row), we show the interactions between the dark sector and the SM particles that could potentially bring them into kinetic equilibrium after the PT. 
The glueballs' heat exchange with SM goes through dark baryons, whose abundance is significantly suppressed during the PT. 
As a result, the glueballs do not maintain equilibrium with the rest of the bath after the PT. 
Instead of calculating the full dark glueball-baryon energy exchange rate, we will simply show that the typical glueball scatters less than once per Hubble time, and so thermalization would not occur even if an $\mathcal{O}(1)$ fraction of the particles' energy was transferred per scattering. We expect that for very heavy dark baryons the energy transferred per scattering will be parametrically small, further suppressing thermalization.

Keeping in mind that the dark baryons eventually make up the DM today, we can write $n_{\mathrm{B}}$ as 
\begin{equation}
    n_\mathrm{B} \sim \frac{10^{-12}}{\xi} \times \left(\frac{1~\mathrm{TeV}}{m_q} \right) \times  T^3,
    \label{eq:eqbGBBnB}
\end{equation}
where $\xi \leqslant 1$ is any further suppression in baryons number density between the PT and today. Thus, at $T=\Lambda$ the scattering rate per Hubble can be rewritten as
\begin{equation}
    \frac{n_{\mathrm{B}} \langle  \sigma v \rangle }{H(T)} \sim  \frac{10^{-1}}{\xi} \times \alpha_{\mathrm{GB}} \alpha_{\mathrm{GB-B}} \times  \frac{\left(1~\mathrm{TeV}\right)^2}{m_\mathrm{DM} \Lambda} ,
    \label{eq:eqbGBBfin}
\end{equation}
where $\alpha_{\mathrm{GB}}$ ($\alpha_{\mathrm{GB-B}}$) is the coupling between glueballs (glueballs and baryons) and is at most $\mathcal{O}(1)$. 
In the upcoming sections we show that $\xi \gtrsim 10^{-4}$ is possible in our setup and we focus on $m_{\mathrm{DM}} \gtrsim 10^4$~TeV and $\Lambda \gtrsim 10$~TeV; 
for this range of variables Eq.~\eqref{eq:eqbGBBfin} indicates that after the PT glueballs and SM particles are not in kinetic equilibrium.

During the PT, the gluon and the glueball gases coexist. The former can transfer its energy and entropy to either the latter or to the SM bath. The expansion of the universe allows the SM and the glueball baths to absorb the released latent heat while their temperature remains roughly constant. In App.~\ref{sec:appxTherm} we provide an argument that this implies the PT takes around a few percent of the Hubble time, which is in agreement with the numerical results of Ref.~\cite{Asadi:2021pwo}.

An important quantity for the rest of this work is the relative entropy density stored in the glueball gas after the PT
\begin{equation}
    R \equiv \frac{s_{\mathrm{GB}}}{s_{\mathrm{SM}}},
    \label{eq:defR}
\end{equation}
where $s_{\mathrm{GB}}$ is the glueball and $s_{\mathrm{SM}}$ the SM entropy density. This ratio remains constant after the PT since the glueballs and SM are not in kinetic equilibrium. In what follows we discuss our prediction for the viable range of $R$.

If the glueball number density after the PT is high enough, glueballs can still maintain  kinetic and chemical equilibrium amongst themselves. 
The chemical equilibrium can be maintained via $3\rightarrow 2$ interactions with large cross sections \cite{PhysRevLett.103.153201,Forestell:2016qhc,Bhatia:2020itt}.

The $3\rightarrow 2$ glueball interaction being in equilibrium suggests that their chemical potential is zero. Furthermore, if the number density is high enough to maintain the glueball gas in kinetic equilibrium with itself, the highest physically meaningful temperature it can have is $T \approx \Lambda$ (for higher temperatures the glueballs melt into gluons). 
This temperature can be used to find an upper bound on the entropy of the glueballs. 
Assuming an equilibrium distribution for $0^{++}$ gas with $m_{0^{++}}\approx 7 \Lambda$ at $T=\Lambda$ and comparing it to the SM bath entropy density at the same temperature\footnote{In doing so we are ignoring possible percent level deviations from $T=\Lambda$ for both the glueball bath and the SM bath.}, we find
\begin{equation}
    R_{\mathrm{\max}} \sim 2.5 \times 10^{-4}.
    \label{eq:Rmax}
\end{equation}
Note that this upper bound was derived with the assumption that enough glueballs form after the PT to maintain their kinetic and chemical equilibrium. 
Accurately calculating the value of $R$ would require detailed numerical studies and is beyond the scope of this work.

\begin{figure*}
    \centering
    \resizebox{1.7\columnwidth}{!}{
    \includegraphics{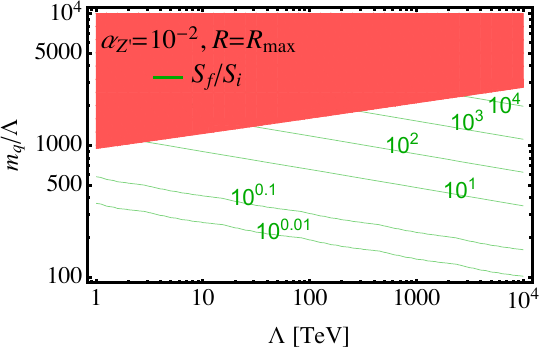}
    \includegraphics{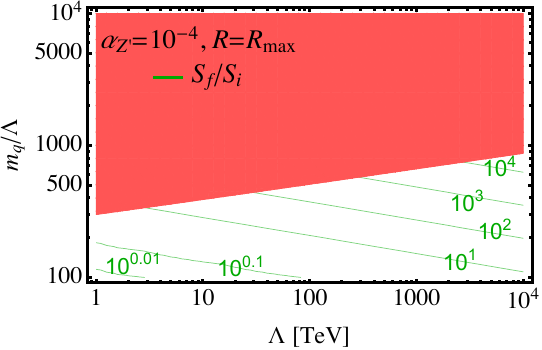}
    }\\
    \resizebox{1.7\columnwidth}{!}{
    \includegraphics{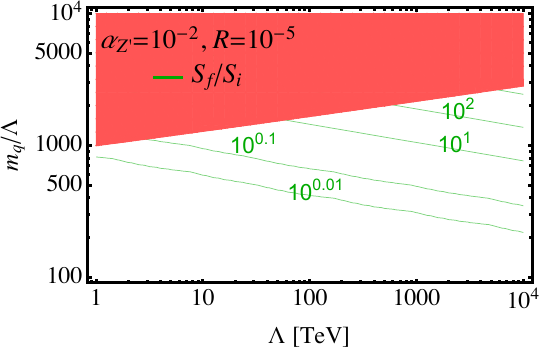}
    \includegraphics{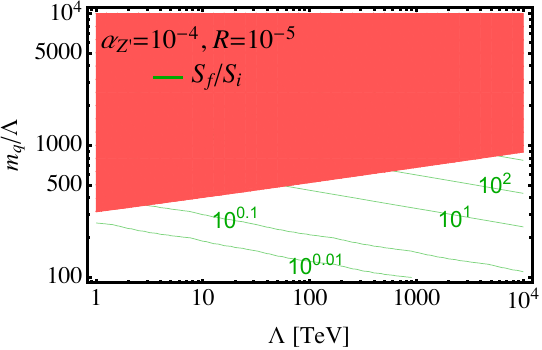}
    }
    \caption{Contours of the ratio of SM entropy after ($S_f$) and before ($S_i$) the glueball decay, for $\alpha_{Z'}=10^{-2}$ (left column), $\alpha_{Z'}=10^{-4}$ (right columns), $R=R_{\mathrm{max}}$ (top row), and $R=10^{-5}$ (bottom row). As $\alpha_{Z'}$ increases, the lifetime of the glueballs ($\tau_{0^{++}}$) decreases, which in turn leads to smaller entropy injection for every point in the parameter space. The contours show substantial entropy injection over a large part of the parameter space. Every point in the red region is ruled out by constraints on the amount of glueballs at the onset of BBN ($t=0.01$s), see the text for more details. }
    \label{fig:Sratioplotfixalpha}
\end{figure*}

In App.~\ref{sec:appxDyn} we provide further dynamical arguments that imply this bound is actually saturated during the PT.  
The analytic estimate we provide in App.~\ref{sec:appxDyn} suggests that the resulting glueball number density is of the order 
\begin{align}
\label{eq:nGBPT}
n_{\rm GB} \approx   2 \cdot 10^{-6} \, T_c^3 \left( \frac{M_{\rm pl}}{T_c} \right)^{0.2}\,,
\end{align}
which implies that the glueball number density is above the value expected from the thermal abundance (the number density corresponding to $R_{\mathrm{max}}$) at $T \approx T_c$ for $T_c < 500\, \text{ TeV}$ and varies very slowly with $T_c$. 
For the range of $T_c$  studied in this work (see upcoming Figs.~\ref{fig:Sratioplotfixalpha}-\ref{fig:finalRLambda}, this approximation suggests that number of glueballs formed after the PT nearly saturates the bound in Eq.~\eqref{eq:Rmax}. 

In addition, we investigate whether the number changing interactions are in equilibrium given the above conditions. Below the confinement temperature, the glueball interactions are described by an effective Lagrangian, a picture confirmed by recent lattice studies~\cite{Yamanaka:2019yek,Yamanaka:2019aeq}. As discussed in~\cite{Forestell:2016qhc}, the cross section for the glueball $3\rightarrow 2$ interaction is given by\footnote{See also Refs.~\cite{Yamanaka:2019aeq,Yamanaka:2019yek}. } 
\begin{align}
    \langle \sigma_{3\rightarrow 2 } v^2\rangle  \approx \frac{1}{(4\pi)^3} \left(\frac{4 \pi}{N_c}\right)^6 \frac{1}{m_{\rm GB}^5} \,.
\end{align}
Using the freezeout condition for this interaction
\begin{align}
   \left( n_{\rm GB}^{\rm min}\right)^2  \langle \sigma_{3\rightarrow 2 } v^2\rangle  = H_{\rm PT}\approx T_c^2/M_{\rm pl}\,,
\end{align}
we find the threshold glueball abundance
\begin{align}
    n_{\rm GB, \rm eq.}^{3 \rightarrow 2} \geq 10^2 \,T_c^3\, \sqrt{\frac{T_c}{M_{\rm pl}} } \,,
\end{align}
above which $3\rightarrow 2$ interactions are in equilibrium. Comparing with Eq.~\eqref{eq:nGBPT}, we find that after the PT this equilibrium condition is satisfied for $T_c < 10^5 \text{ TeV}$. The parameter space studied in this work is within this range, see the upcoming Figs.~\ref{fig:Sratioplotfixalpha}-\ref{fig:finalRLambda}. Therefore, in the parameter region we will investigate, those number changing processes will promptly reduce the number density of glueballs right after the PT to the thermally predicted value, saturating $ R_{\mathrm{max}}$.

All in all, we see that even if we start from $R \geqslant R_{\mathrm{max}}$ during the PT, $3\rightarrow 2$ freeze-out interactions bring $R$ down to $R_{\mathrm{max}}$ immediately after the PT. These glueballs are then decoupled from the SM bath and their entropy is conserved. While the analysis of App.~\ref{sec:appxDyn} suggests $R \sim R_{\mathrm{max}}$ after the PT, it overlooks many complications of the PT and needs to be corroborated in explicit numerical simulations. 
As a result, we will use $R\leqslant R_{\mathrm{max}}$ as an input in our calculation. 

We use the analytic technique developed in Ref.~\cite{Forestell:2016qhc} to calculate the asymptotic glueball abundance after the PT; 
the error introduced by using the analytic approximation of Ref.~\cite{Forestell:2016qhc} (as opposed to the full numerical treatment) is subdominant to the uncertainty introduced by the PT dynamics; see Ref.~\cite{Asadi:2021pwo} for further details.

We assume that $0^{++}$, the lightest glueball, is the only one that survives after the PT (see App.~\ref{sec:appxGB} for an explanation why this is a good approximation). 
Following the calculation of Ref.~\cite{Forestell:2016qhc}, one can show that 
\begin{equation}
    x_{0^{++},fo}^{5/2} e^{2x_{0^{++},fo}} \approx \frac{g_{*S,fo}}{180\pi} R \left( \frac{m_{0^{++}}^4 M_{pl} \langle \sigma_{32} v^2 \rangle }{\sqrt{ \frac{4}{5} \pi^3 g_{*,fo}}} \right)^{3/2}
    \label{eq:Rxfo}
\end{equation}
where $x_{0^{++}}=m_{0^{++}}/T_{0^{++}}$ with $m_{0^{++}} \approx 7 \Lambda$ \cite{Morningstar:1999rf,Mathieu:2008me} ($T_{0^{++}}$) being the lightest glueball mass (temperature of the glueball bath), 
$\langle \sigma_{32} v^2 \rangle$ is the interaction rate for the $3\rightarrow 2$ number-changing process that controls glueballs' abundance after the PT, 
subscript $fo$ refers to the time of glueball freezeout after the PT, and $g_{*(S),fo}$ is the relativistic degrees of freedom (for entropy) when the glueball number-changing process freezes out. We can solve this equation iteratively to find $x_{0^{++},fo}$. The asymptotic glueball abundance then can be written as \cite{Forestell:2016qhc}
\begin{equation}
    Y_{0^{++}} \approx \frac{R}{x_{0^{++},fo}}.
    \label{eq:YGBRx}
\end{equation}

In this calculation we are neglecting the finite glueball lifetime. 
For large enough values of $R$, glueballs can eventually dominate the energy budget of the universe before they decay, giving rise to an early matter-dominant epoch. 
Using Eq.~\eqref{eq:YGBRx} we can calculate the energy density of the glueballs after their freezeout, which in turn can be used to calculate the temperature $T_i$ at which the universe enters an early matter-dominant epoch. 
Assuming $g_{*}=g_{*S}$ at $T=T_i$ (which is a good approximation since the only particle with different temperature than SM is the non-relativistic glueball), we can show
\begin{equation}
    T_i = \frac{4}{3} m_{0^{++}} Y_{0^{++}}.
    \label{eq:Ticalc}
\end{equation}
Combining Eqs.~\eqref{eq:Rxfo}-\eqref{eq:Ticalc} we can calculate $T_i$ as a function of $R$.

Eventually, these glueballs have to decay away to SM before BBN, see for instance Refs.~\cite{Jedamzik:2006xz,Kawasaki:2017bqm}. In our model, this decay is facilitated via the interaction with a $Z'$ and proceeds through a dark quark loop (see App.~\ref{sec:appxGB} for further details of various glueballs decay rates).

\begin{figure}
    \centering
    \resizebox{0.8\columnwidth}{!}{
    \includegraphics{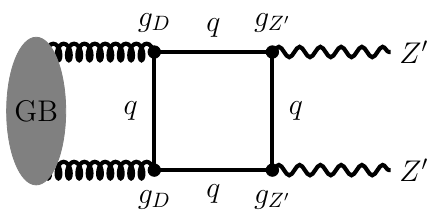}
    }
    \caption{ The 1-loop diagram giving rise to the $0^{++}$ glueball decay. For the range of masses we study, the final $Z'$s are on-shell. The heavy quark loop gives rise to a suppression of $\left(\Lambda/m_q\right)^8$ in the glueball's lifetime \cite{Juknevich:2009ji,Juknevich:2009gg}; for the heavy quarks in our model this can naturally give rise to long-lived glueballs.}
    \label{fig:GBdecay}
\end{figure}

The decay channel of $0^{++}$ glueballs is shown in Fig.~\ref{fig:GBdecay}. For $m_{Z'} \leqslant \Lambda$, the $Z'$ mass gives rise to less than $10\%$ correction in $0^{++}$ lifetime; thus for simplicity we neglect the effect of $m_{Z'}$ and find the following simple formula for the glueballs decay rate
\begin{equation}
    \Gamma = \frac{\alpha_D^2 \alpha_{Z'}^2}{4\pi} \left( \frac{1}{60}\right)^2 \frac{m_{0^{++}}^3 F_{0^{++}}^2}{m_q^8},
    \label{eq:decayformula}
\end{equation}
where $\alpha_D$ ($\alpha_{Z'}$) is the dark confining gauge symmetry's (new $\ubl$'s) structure constant, $m_q$ is the dark quark mass, and $F_{0^{++}}$ is the glueball's decay constant; 
we use \cite{Chen:2005mg,Juknevich:2009gg}
\begin{equation}
    F_{0^{++}} \approx \frac{m_{0^{++}}^3}{4\alpha_D (m_{0^{++}})},
    \label{eq:F0++}
\end{equation}
with $\alpha_D (m_{0^{++}})$ denoting the dark structure constant evaluated at the mass of the lightest glueball.

Note that for $m_{Z'} \geqslant m_{0^{++}}/2$ the decay channel of Fig.~\ref{fig:GBdecay} is forbidden and $0^{++}$ glueballs can only decay to SM via off-shell $Z'$ particles. By focusing on the range $m_{Z'} \leqslant \Lambda$ we avoid this extra complication.

\subsection{Entropy injection}
\label{subsec:entropy}

The glueballs' energy density can be calculated as a function of $R$ using Eqs.~\eqref{eq:Rxfo}-\eqref{eq:Ticalc}. They will eventually decay to SM particles at $T=T_{\tau_{0^{++}}}$, injecting a substantial amount of entropy into the thermal bath that dilutes the DM abundance further. 
The calculation for the amount of this entropy injection and the dilution it gives rise to can be found in the literature, e.g. see Refs.~\cite{Kolb:1990vq,Berlin:2016gtr,Contino:2018crt,Asadi:2021bxp}. In the case of a single unstable glueball the result is 
\begin{equation}
    \xi \equiv \left(\frac{S_f}{S_i}\right)^{-1}  = \left( 1+ 1.65 <g_{*S}^{1/3}> \left(\frac{T_i^4}{(\Gamma M_{pl} )^{2}} \right)^{1/3} \right)^{-3/4}.
    \label{eq:Entinjection}
\end{equation}
where $S_f$ ($S_i$) is the SM bath total entropy after the decay (before the decay), $T_i$ is the temperature at which the universe enters the early matter-dominant epoch, $\Gamma$ is the decay rate of the unstable particle, and $\langle g_*^{1/3} \rangle$ is an average of the relativistic degrees of freedom in the early universe (defined in the appendix of Ref.~\cite{Asadi:2021bxp}).

Equation~\eqref{eq:Entinjection} clearly shows that the entropy injection is larger when the glueballs have a larger lifetime, i.e. smaller decay rate $\Gamma$.
There are, however, bounds on the lifetime of a long-lived relic in the early universe from BBN, see for instance Refs.~\cite{Jedamzik:2006xz,Kawasaki:2017bqm}. For simplicity, in this work we use the bound of 
\begin{equation}
    \Omega_{0^{++}} \leqslant 10^3 ~~ \mathrm{at} ~ t=0.1~\mathrm{s},~ \& ~ \tau_{0^{++}} \leqslant 10^{-2}~\mathrm{s}
    \label{eq:BBN}
\end{equation}
(see App.~\ref{sec:appxGB} for more details). This puts a bound on the amount of entropy injection and DM dilution that can take place in the early universe.

In Fig.~\ref{fig:Sratioplotfixalpha} we show the contours of $S_f/S_i$ on a plane of $\Lambda-m_q/\Lambda$ for two different values of $\alpha_{Z'}$ and two different values of $R$. 
It is evident that we can indeed have non-negligible entropy injection for a large part of the parameter space.

To understand the shape of the contours in Fig.~\ref{fig:Sratioplotfixalpha} we should keep in mind that for a fixed $\alpha_{Z'}$, as we go to higher $m_q/\Lambda$ ratios, Eq.~\eqref{eq:decayformula} suggests a smaller $\Gamma$, which in turns suggests more entropy injection (see Eq.~\eqref{eq:Entinjection}).  
Furthermore, going to larger $\Lambda$ (with fixed $m_q/\Lambda$) means $T_i$  increases, which again according to Eq.~\eqref{eq:Entinjection} implies larger entropy injection as well.

Comparing the two columns of Fig.~\ref{fig:Sratioplotfixalpha} we see that when the gauge coupling decreases, a smaller part of the parameter space gives rise to glueball decays fast enough to avoid the BBN bounds. 
On the other hand, this prolonged lifetime means a larger entropy injection for a fixed point on the $\Lambda-m_q/\Lambda$ plane. The figures also suggest that larger values of $R$ imply a larger entropy injection for a fixed point in the parameter space, as expected.

\begin{figure}
    \centering
    \resizebox{1\columnwidth}{!}{
    \includegraphics{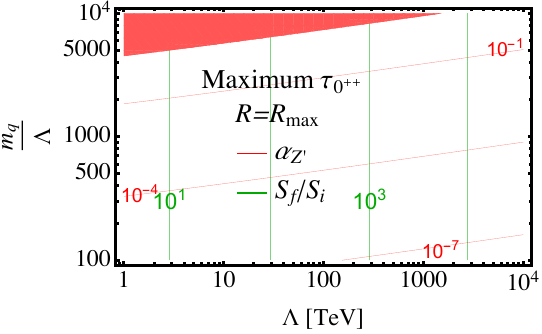}
    }\\
    \resizebox{1\columnwidth}{!}{
    \includegraphics{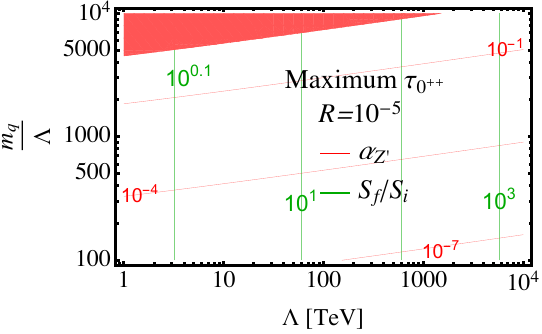}
    }
    \caption{Contours of maximum $S_f/S_i$ (green) and the required $\alpha_{Z'}$ (red) for $R=R_{\mathrm{max}}$ from Eq.~\eqref{eq:Rmax} (top) and $R=10^{-5}$ (bottom). For each point on the plane $\alpha_{Z'}$ is set to the value that maximizes the glueball lifetime $\tau_{0^{++}}$ given the BBN constraints discussed in the text. In the red region the required $\alpha_{Z'} > \sqrt{4\pi}$ and so our perturbative treatment breaks down. We find substantial entropy injection that can further dilute the DM abundance for most of the parameter space.}
    \label{fig:Sratioplotmax}
\end{figure}

If instead of fixing $\alpha_{Z'}$, we fix the glueball lifetime to the maximum value allowed by BBN bounds~\cite{Jedamzik:2006xz,Kawasaki:2017bqm}, we get the maximum amount of entropy injection for any point in the parameter space. 
This condition can be used to calculate $\alpha_{Z'}$ for every point on the parameter space. 
In the rest of this work we focus on this ``maximum dilution" scenario to find the upper bound on how much the glueball entropy injection can push the relic abundance line to higher masses. 
The contours of constant entropy injection ratio $S_f/S_i$ in this scenario are shown in Fig.~\ref{fig:Sratioplotmax} for two different values of $R$.

The $m_q/\Lambda$ ratio enters the entropy injection calculation only via $\Gamma$, see Eq.~\eqref{eq:Entinjection}. Thus, since in Fig.~\ref{fig:Sratioplotmax} we fix $\Gamma$ for any given $\Lambda$, the $S_f/S_i$ contours are independent of $m_q/\Lambda$. Notice that to keep $\Gamma$ fixed as we increase $m_q/\Lambda$, $\alpha_{Z'}$ increases as well. 
As a result, eventually for large enough values of $m_q/\Lambda$ the required $\alpha_{Z'}$ exceeds $\sqrt{4\pi}$ and our perturbative treatment breaks down. This puts an upper bound on $m_q/\Lambda$, denoted by the red region on the plane of Fig.~\ref{fig:Sratioplotmax}. 
The value of $\alpha_{Z'}$ everywhere on Fig.~\ref{fig:Sratioplotmax} is large enough to guarantee $Z'$ is in kinetic equilibrium with SM before it decays.

Due to the entropy injection, DM can have masses even higher than predicted by the squeezeout effect of Refs.~\cite{Asadi:2021yml,Asadi:2021pwo}. 
In the upcoming section we combine the suppression factors from the squeezeout during the PT and the glueball decay to find the relic abundance line on our parameter space.
As mentioned earlier, in our calculation we used the central prediction of Refs.~\cite{Asadi:2021yml,Asadi:2021pwo} for the pocket survival factor during the squeezeout; various sources of uncertainty in modeling of the PT in Refs.~\cite{Asadi:2021yml,Asadi:2021pwo} give rise to around an order of magnitude uncertainty in the DM mass in each direction in the upcoming results.

\section{The Phenomenology}
\label{sec:pheno}

In the previous section we found that the glueball decay in our model can significantly change the viable parameter space compared to Refs.~\cite{Asadi:2021yml,Asadi:2021pwo}. 
In this section we study the signals of our model in various current experiments. 
To avoid cluttering the discussion, we only focus on the scenario where the glueball lifetime $\tau_{0^{++}}$ is set to its maximum allowed by BBN, see Fig.~\ref{fig:Sratioplotmax} and the discussions around it. 
In this scenario the DM mass is pushed to the highest values possible in our setup. 
We will find that while some parts of the parameter space are already probed and ruled out, there is still viable parameter space left that can be probed in future experiments.

\begin{figure*}
    \centering
    \resizebox{2\columnwidth}{!}{
    \includegraphics{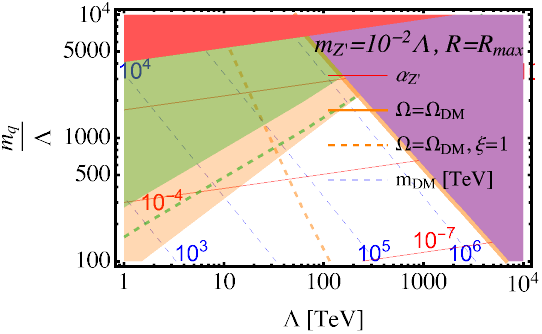}
    \includegraphics{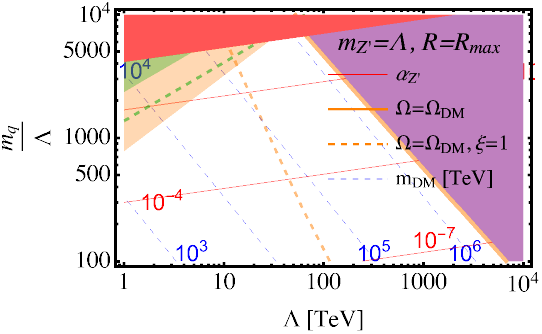}
    }\\
    \resizebox{2\columnwidth}{!}{
    \includegraphics{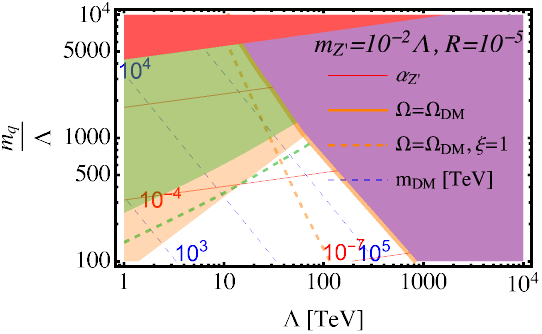}
    \includegraphics{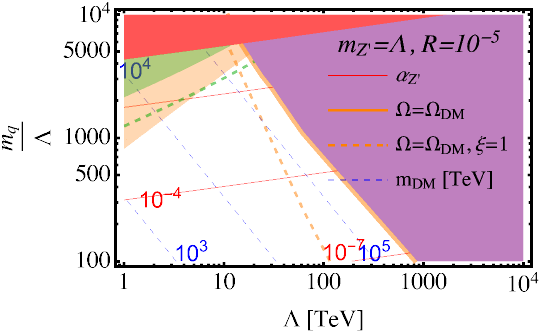}
    }
    \caption{The relic abundance line (solid orange), and constraints from direct detection (green) and collider (orange) searches on the parameter space of our model for different values of $m_{Z'}$ and $R$. 
    For any point below the dashed green line the scattering rate at the direct detection experiments will be below the neutrino floor. 
    Contours of constant DM mass are denoted by dashed blue lines. 
    In each plot the purple region corresponds to where the universe is overclosed. 
    The relic abundance line when the glueball decay is neglected is denoted by the dashed orange line. 
    We set $\alpha_{Z'}$ (red lines) to the value that minimizes $\Gamma$ as allowed by the BBN bounds (similar to Fig.~\ref{fig:Sratioplotmax}). 
    The red region is ruled out since the required $\alpha_{Z'}$ there exceeds $\sqrt{4\pi}$ and so we expect perturbatively calculated rates to break down.  
    Depending on the $Z'$ mass, bounds on the model can change significantly. The value of $R$ can substantially affect the final relic abundance line location; for lower values of $R$ the relic abundance line approaches the dashed orange line. 
    While a large part of the parameter space is already ruled out, viable parameter space still exists. }
    \label{fig:moneyplotmax}
\end{figure*}

There are five free parameters in our setup ($m_q$, $\Lambda$, $\alpha_{Z'}$, $m_{Z'}$, and $R$) of which one ($\alpha_{Z'}$) is fixed by the maximum glueball lifetime assumption for every point in the parameter space. 
In Fig.~\ref{fig:moneyplotmax}, the bounds from different experiments are summarized for different values of $m_{Z'}$ and $R$ on a plane of $\Lambda - m_q/\Lambda$. 
In what follows we review different searches that can probe our model and explain different regions in the figure.

Before that, we should note that Fig.~\ref{fig:moneyplotmax} suggests that as the value of $R$ increases, the relic abundance line gets pushed to higher DM masses, since more entropy will be injected into the SM after the glueballs decay. For low enough $R$ values the injected entropy can be neglected and our prediction for the relic abundance line goes back to the prediction of Ref.~\cite{Asadi:2021yml,Asadi:2021pwo} (the dashed orange line in Fig.~\ref{fig:moneyplotmax}). The case of $R=R_{\mathrm{max}}$ corresponds to the largest amount of entropy injected, and thus the highest DM mass consistent with today's observed DM abundance.

\subsection{Direct detection}
\label{subsec:DD}

Our DM particles scatter elastically and coherently off  nuclei via a t-channel exchange of $Z'$. Direct detection experiments can probe this scattering rate and put a bound on the DM-nucleon scattering rate. The cross section for this process can be calculated straightforwardly
\begin{equation}
    \sigma_{\mathrm{DD}} \approx 144 \pi \alpha_{Z'}^2 \frac{\mu_N^2}{m_{Z'}^4},
    \label{eq:DDrate}
\end{equation}
where $\mu_N = m_{\mathrm{DM}} m_\mathrm{N}/(m_{\mathrm{DM}} + m_\mathrm{N})$ is the reduced mass of the incoming DM particle and the target nucleon with mass $m_\mathrm{N}$. 
In writing this equation we have assumed the transferred momentum $q \sim \mu_N v_{\mathrm{DM}}$ (where $v_{\mathrm{DM}}$ is the incoming DM velocity) is much smaller than $m_{Z'}$; we can check that for the allowed $Z'$ masses due to the collider constraints (discussed in the next section), this is always satisfied.

The rate in Eq.~\eqref{eq:DDrate} should be compared to the bounds on the DM-nucleon cross section. 
In Fig.~\ref{fig:moneyplotmax} we show the current bounds from the XENON1T experiment \cite{XENON:2018voc} (green regions), as well as the contour with the same rate as the neutrino floor in the future XENONnT experiment \cite{XENON:2020kmp}. 
For any point below the neutrino floor contour (dashed green) the neutrino scattering gives rise to a large background and novel search techniques are needed~\cite{Hochberg:2016ntt,Essig:2016crl,Kadribasic:2017obi,Budnik:2017sbu,Rajendran:2017ynw,Griffin:2018bjn,Coskuner:2019odd,Blanco:2021hlm}, see also Refs.~\cite{Ebadi:2022axg,Akerib:2022ort} for recent reviews.

Depending on $R$ and $m_{Z'}$, we find that direct detection experiments can probe part of the relic abundance line in our model.

Note that according to Eq.~\eqref{eq:DDrate}, since $\mu_N$ goes to $m_\mathrm{N}$ in the limit of $m_{\mathrm{DM}} \gg m_\mathrm{N}$, the rate will be independent of the DM mass. As we go to higher $\Lambda$, i.e. higher $m_{Z'}$, the scattering rate is suppressed and we can only probe larger values of $\alpha_{Z'}$. This explains the shape of green regions in Fig.~\ref{fig:moneyplotmax}.

As discussed earlier, we only consider $m_{Z'} \leqslant \Lambda$. From Fig.~\ref{fig:moneyplotmax} it is clear that if we had gone to higher values of the $Z'$ mass the direct detection experiments would lose their sensitivity.

\subsection{Indirect detection}
\label{subsec:ID}

Direct annihilation of dark sector baryons is highly suppressed~\cite{Witten:1979kh,Geller:2018biy}, and takes place in a rearrangement reaction, where the quarks in the baryon and anti-baryons are arranged to form unstable meson states that subsequently decay to SM particles~\cite{Dondi:2019olm}.

The capture cross section is dominated by the large angular momentum modes, that are limited by $\ell \approx R_B\times p$, where $R_B = (\alpha_D (m_q) \, m_q)^{-1}$ is the baryon Bohr radius and $p$ its momentum. As found in Refs.~\cite{Mitridate:2017oky,Geller:2018biy} the cross section is close to the geometric bound (with an additional suppression factor at large kinetic energies $E_{\rm kin}$) and is given by
\begin{eqnarray}
    \label{eq:IDeq}
    \langle \sigma_{\rm B \bar{B}}  v_{\rm rel} \rangle &=& \frac{\pi R_B^2 \,    \langle v_{\rm rel} \rangle /2 }{\sqrt{E_{\rm kin}/E_B}} = \frac{\sqrt{3}\pi}{4 \alpha_{D} m_q^2}\\ 
    &\sim & 10^{-30} \frac{\text{cm}^3}{\text{s}} \, \left( \frac{10^4 \, \text{TeV}}{m_{\mathrm{DM}}}\right)^2 \left( \frac{0.1}{\alpha_D}\right) \,, \nonumber
\end{eqnarray}
where $v_{\rm rel}$ is the dark baryons' relative velocity, and $E_B$ their binding energy.

In principle residual interactions between the baryons can lead to a Sommerfeld enhancement \cite{Hisano:2004ds,Hisano:2006nn,Cirelli:2007xd,Arkani-Hamed:2008hhe,Cassel:2009wt,Slatyer:2009vg} of the annihilation. 
In the regime of $\Lambda/m_q \ll 1$, the baryon coupling to glueballs is very small, preventing a large Sommerfeld enhancement. 
However, for small enough $m_{Z'}$, interactions with $Z'$ give rise to an enhancement of the annihilation rate in Eq.~\eqref{eq:IDeq} by $\alpha_{Z'}/v_{\rm rel} \lesssim 10^{2} - 10^{3}$. 

Nonetheless, the indirect detection experiments such as Fermi-LAT \cite{Fermi-LAT:2017opo} or IceCube \cite{IceCube:2017rdn}
are not sensitive enough to test cross sections below $10^{-26}  \text{cm}^3/\text{s}$ at such high energies. 
Thus, even if Sommerfeld enhancement takes place in a concrete model, the sensitivity of present day experiments is insufficient to probe our viable parameter space.

However, there is still some potential for interesting indirect-detection signals, requiring further study.
The early matter-dominant epoch in our setup gives rise to the formation of dense DM substructures at small scales. These dense clumps could boost the DM annihilation rate today \cite{Erickcek:2011us,StenDelos:2019xdk}. 
Furthermore, excitation and de-excitation of dark baryons in dense regions could enhance the signal in indirect detection experiments. 
DM particles could also accumulate in the potential well of astrophysical objects today. This can enhance indirect detection signal or affect the evolution of such bodies, e.g. see Refs.~\cite{PhysRevD.40.3221,Leane:2017vag,Baryakhtar:2017dbj,Leane:2021ihh}. 
We leave a detailed study of these effects for future works.

\subsection{Collider bounds}
\label{subsec:collider}

Since DM candidates in our model are heavier than the unitarity bound, it is impossible to produce them at collider facilities in the foreseeable future. However, the $Z'$ portal can potentially have signals at the LHC. There are various searches looking for a heavy $Z'$ beyond the SM at the LHC. In Fig.~\ref{fig:moneyplotmax} we use the most recent non-resonance searches at ATLAS \cite{ATLAS:2020yat} and CMS \cite{CMS:2021ctt} to derive the collider bounds on the model. These searches study the interference with SM Drell-Yan process with final electrons or muons. The bounds from these models can be approximated as
\begin{equation}
    \frac{m_{Z'}}{\sqrt{\alpha_{Z'}} } \gsim \mathcal{B} ,
    \label{eq:collider1}
\end{equation}
with $\mathcal{B}$ read from the results of Refs.~\cite{ATLAS:2020yat,CMS:2021ctt}.
The exact number used on the right-hand side can change depending on the the final state ($ee$, $\mu\mu$, or $ee + \mu\mu$) and is at most in the $\mathcal{B} \approx 30$ TeV ballpark:
\begin{equation}
    m_{Z'} \gsim 30 \sqrt{\alpha_{Z'}} ~ \mathrm{TeV} .
    \label{eq:collider2}
\end{equation}

The value of $\alpha_{Z'}$ in Fig.~\ref{fig:moneyplotmax} is set from the calculation of the glueballs' lifetime. 
In Fig.~\ref{fig:moneyplotmax} we use different values of $m_{Z'}/\Lambda$ in each plot and use Eq.~\eqref{eq:collider2} to put bounds on the model. 
To explain the shape of the excluded region (orange region), note that in the figure as we go to lower values of $\Lambda$ the $Z'$ mass decreases as well, while as we go to higher $m_q/\Lambda$ values $\alpha_{Z'}$ increases.

\subsection{Other signals}
\label{subsec:othersignals}

Our setup can potentially give rise to many other signals in different experiments, see Ref.~\cite{Asadi:2021pwo}. Nonetheless, many of these signals are suppressed in the model and range of parameters studied here. Below we briefly comment on a few such suppressed signals that can not be detected in the foreseeable future.

The first order confinement PT in our model can give rise to gravitational waves (GWs). 
It should be noted that the GW spectrum is independent of the heavy quark mass as the quarks are decoupled by the time of the PT and do not affect it. 
The GW signals from PTs are a function of a few thermodynamic parameters and have been studied extensively in the literature, e.g. see Refs.~\cite{Flanagan:2005yc,Schwaller:2015tja,Caprini:2015zlo,Jaeckel:2016jlh,Katz:2016adq,Bartolo:2016ami,Weir:2017wfa,Croon:2018erz,Ellis:2018mja,Breitbach:2018ddu,Helmboldt:2019pan,Caprini:2019egz,Brower:2020mab,Huang:2020mso}. 

One of these thermodynamic quantities is the velocity of the bubbles during the percolation. Unfortunately, for the non-relativistic velocities in our model of the PT the signals will be strongly suppressed and will not be detected by any of the proposed detectors in the foreseeable future \cite{Caprini:2015zlo}.

Furthermore, the residual dark QCD interactions give rise to some self-interaction among DM particles. Nonetheless, for the heavy DM candidates in our setup we can easily show that the self-interaction rate is far below the current upper bounds \cite{Markevitch:2003at,Feng:2009mn,Buckley:2009in,10.1111/j.1365-2966.2012.21182.x,10.1093/mnrasl/sls053,Tulin:2017ara,Bondarenko:2020mpf} and do not give rise to any detectable signals.

The squeezeout mechanism opened up parameter space above the unitarity bound on thermal DM models through inhomogeneous DM distribution during the PT. 
In Ref.~\cite{Asadi:2021pwo} the possibility of detecting such inhomogeneities was studied for the range of $\Lambda$ studied here and it was shown that this effect is too small to be detected in on-going surveys.

Another intriguing scenario arises in the case that the $Z'$ mass becomes small, which leads to long range interactions. This can substantially increase the capture probability of the dark baryons in compact objects, such as exoplanets~\cite{Leane:2020wob}, leading either to heat signatures in the case of prompt $Z'$ decay in the objects, or to new high energy signals at the surface of those objects~\cite{Leane:2021ihh,Leane:2021tjj}. 
A detailed study of this signal is outside the scope of this work and is left for future studies.

Finally, the possibility of forming stable macroscopic objects has been studied in similar setups \cite{Gross:2021qgx} as well. However, it has been shown that such objects can only form when both $m_q/\Lambda$ and $m_q$ are much larger than the range considered in our setup.

\begin{figure*}
    \centering
    \resizebox{2\columnwidth}{!}{
    \includegraphics{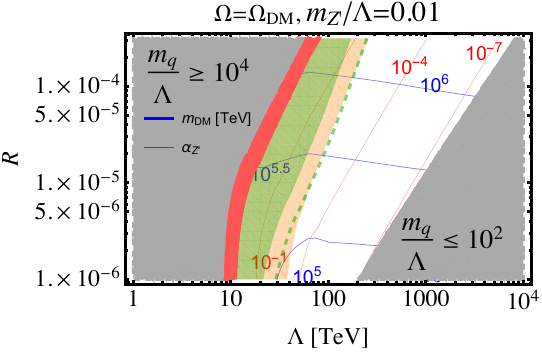}
    \includegraphics{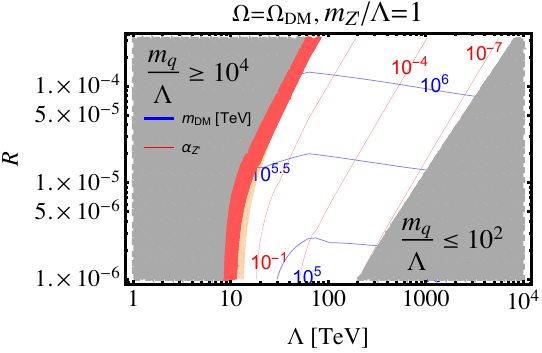}
    }
    \caption{The available parameter space of the model. We use the maximum dilution assumption and fix the relic abundance to the observed value to calculate $\alpha_{Z'}$ and dark matter mass as a function of $\Lambda$ and $R$. 
    Contours of $\alpha_{Z'}$ (red) and dark matter mass (blue) are shown. In the gray region we fall outside the $m_q/\Lambda$ window where the analysis of Ref.~\cite{Asadi:2021pwo} is viable. 
    The red region denotes where values of $\alpha_{Z'}$ required by the maximum dilution assumption exceed $\sqrt{4\pi}$ and so the assumption of perturbativity is violated. 
    Bounds from direct detection (green) and collider searches (orange) are shown as well, see the previous sections for further details; the dashed green line denotes the boundary of neutrino floor in direct detection searches; novel search techniques are required to probe the parameter space to the right on this line in direct detection experiments. 
    These bounds can change as $m_{Z'}$ varies. 
    We find a large available parameter space. 
    Depending on the value of $R$, whose precise calculation would require further extensive numerical work including lattice simulations, we can get the right DM abundance with DM masses roughly in the $\left( 10^5,~10^6 \right)$~TeV range.   }
    \label{fig:finalRLambda}
\end{figure*}

\subsection{Summary of signals}
\label{subsec:signalssummary}

As argued before, we have five free parameters in our setup: $m_q$, $\Lambda$, $m_{Z'}$, and $\alpha_{Z'}$ are the parameters of the model, while $R$ parameterizes our lack of understanding of the details of the PT. With the ``maximum dilution" assumption we can fix the value of $\alpha_{Z'}$ as a function of the other parameters. Furthermore, as evident from Fig.~\ref{fig:moneyplotmax} for any fixed value of $\Lambda$ and $R$ the DM mass (or equivalently dark quark mass $m_q$) can be determined such that we get the right DM relic abundance.

In Fig.~\ref{fig:finalRLambda} we show the viable parameter space on the plane of $\Lambda-R$ for different values of $m_{Z'}$. The maximum dilution assumption and getting the right DM relic abundance are used to calculate $m_q$ and $\alpha_{Z'}$ as a function of $\Lambda$ and $R$. 
Details of the direct detection and collider constraints are discussed in the previous sections. These bounds strongly depend on the value of $m_{Z'}$ as depicted by the figure.

Figure~\ref{fig:finalRLambda} shows that as we go to higher values of $R$ the relic abundance line moves to higher masses; this was expected since $R$ is an indicator of how much entropy is stored in the glueball bath and can be injected into the SM after its decay. 
The highest value of $R$ shown in the figure is $R_{\mathrm{\max}}$ defined in Eq.~\eqref{eq:Rmax}; recall that we derived this bound by assuming enough glueballs are produced such that the glueball $3\rightarrow 2$ process attains equilibrium, and that glueballs then follow a thermal distribution with $T=\Lambda$. 
If we consider lower values of $R$, eventually the injected entropy becomes so small that it can not affect the relic abundance of dark matter and we will recover the prediction of Ref.~\cite{Asadi:2021yml,Asadi:2021pwo} for the DM mass. 

As argued earlier (see also App.~\ref{sec:appxDyn}), there are some arguments suggesting $R_\mathrm{max}$ is saturated after the PT. 
For this value of $R$, Fig.~\ref{fig:finalRLambda} suggests the DM mass is pushed to values as high as $\sim 10^6$~TeV.

\section{Conclusions}
\label{sec:conclusion}

In this work we provided a specific model of ``thermal squeezeout" \cite{Asadi:2021yml,Asadi:2021pwo} with a $Z'$ portal between the dark and the visible sector. 
We pointed out that in the range of parameters where the squeezeout effect becomes relevant, dark glueballs can naturally be long-lived enough that they can give rise to an early matter-dominant epoch and inject significant entropy into the SM, which can further push the viable DM mass to higher values.

The final DM mass range is subject to  uncertainty in calculating the survival factor during the PT \cite{Asadi:2021yml,Asadi:2021pwo}, as well as uncertainty in the amount of entropy in the glueball bath after the PT, captured by the parameter $R$ (see Eq.~\eqref{eq:defR}). These uncertainties arise from our lack of understanding of the details of a confinement PT in the early universe, and motivate future work on the subject. 

All in all, we found that if we neglect the uncertainty in determining the survival factor during the PT, in the most extreme case, DM masses as high as $\sim 10^6$~TeV are viable in this setup. Including the uncertainties in calculating the survival factor (see Ref.~\cite{Asadi:2021pwo} for further details) could potentially push this upper bound to around $\sim 10^7$~TeV.

We identified the parameter space of the model that is already ruled out by various experiments. In particular, direct detection and collider bounds can rule out some parts of the parameter space for sufficiently light  $Z'$ masses. The model still has a large viable parameter space that can be probed in future experiments. The fact that the rich dynamics in minimal confining dark sectors, such as our setup here, can open up parameter space for DM masses far above the often-quoted unitarity bound \cite{Griest:1989wd,vonHarling:2014kha,Smirnov:2019ngs} should motivate developing new searches that can probe this part of the parameter space.

The model can also have further interesting signals whose in-depth study is left for future works. 
Firstly, it could give rise to interesting signals when a DM particle excites to other hadrons and, subsequently, radiates energy during its de-excitation. 
Accumulation of the DM around various celestial bodies could lead to novel indirect detection signals or modify the properties and evolution of the bodies in question \cite{PhysRevD.40.3221,Leane:2017vag,Baryakhtar:2017dbj,Leane:2021ihh}. 
Furthermore, the early matter-dominant epoch in our setup, which is required for significant entropy injection after the glueballs decay, can also give rise to interesting signatures in the matter power spectrum and indirect detection experiments~\cite{Erickcek:2011us,StenDelos:2019xdk,Barenboim:2021swl}, as well as modifying the spectrum of gravitational waves from other sources~\cite{Cui:2018rwi,Auclair:2019wcv,Gouttenoire:2019kij,Hook:2020phx,Chang:2021afa,Ertas:2021xeh}.

\section*{Acknowledgments}

We thank Cari Cesarotti, William Detmold, Yann Gouttenoire, Di Liu, Patrick Meade, Ofri Telem, and Ken Van Tilburg for useful discussions. 
The work of PA and TRS was supported by the U.S. Department of Energy, Office of Science, Office of High Energy Physics, under grant Contract Number DE-SC0012567. EK and EDK are supported by the Israel Science Foundation (grant No. 1111/17) and by the Binational Science
Foundation (grants No. 2020220). JS is supported by the European Research Council under grant Number 742104.

\appendix

\section{Thermodynamic Approach to the Phase Transition}
\label{sec:appxTherm}

In our previous work \cite{Asadi:2021pwo}, a numerical analysis was developed for studying how long the PT takes. Here we provide an estimate of the length of the phase transition based purely on thermodynamic considerations. We will show that, throughout the phase transition, entropy is approximately conserved, except for the initial bubble formation, and for effects that are proportional to the amount of supercooling $\epsilon \equiv (T_c-T)/T_c$. The total entropy production will be given by $\int\!\epsilon(x)\,dx$, where $x$ is the phase fraction converted to confined phase. To prove that the entropy is approximately conserved, we assume that the temperature $1/T \equiv \left({\partial S}/{\partial E}\right)_V$ and that the pressure $P \equiv T\left({\partial S}/{\partial V}\right)_E$ are well-defined during the phase transition. The condition for this to be valid is that local kinetic equilibrium be maintained. This will be the case if the scattering time $t_{\rm sc} \equiv 1/(n\sigma v)$ be much less than the time over which the energy density changes by an order 1 fraction, i.e. the total time for the phase transition, since the latent heat $L$ is an order $10^{-1}$ fraction of the standard model density $\rho_{\rm SM}$. In this case, we can assume the first law of thermodynamics during the phase transition:
\begin{align}
\label{eq:1stlaw}
    dS = \frac{dE}{T} + \frac{P}{T}\,dV\;,
\end{align}
for a given comoving volume $V$. We also note that the the second Friedman equation, describing cooling and dilution due to Hubble expansion,
\begin{align}
\label{eq:friedmann}
    \dot \rho = -3H(\rho + P)
\end{align}
can be written, defining a comoving volume $V=a^3(t)$, in the form
\begin{align}
    \label{eq:pdV}
    dE = - P\, dV
\end{align}
where $E\equiv\rho V$ and $dV = 3 H V dt$. Here, $P$ is the internal pressure of that gas, which means the gas is doing the maximum work allowed by its pressure. Such maximum work processes are special and imply conservation of entropy. 
Plugging Eq.~\eqref{eq:pdV} back into the first law \eqref{eq:1stlaw}, we indeed find that $dS=0$ identically for any comoving volume. 

The question that remains is whether \eqref{eq:pdV} remains valid during the phase transition, when the metric and stress tensor are both inhomogeneous. In fact, we can derive a version of \eqref{eq:pdV} that does not assume homogeneity. Assuming the metric has the form $g_{\mu\nu}={\rm diag}(1,-g^{(3)}_{ij}(\vec{x},t))$, with $g^{(3)}$ the metric of the spatial splices, we can write covariant conservation of stress-energy $\nabla_\mu T^{\mu0}=0$ as: \begin{align}
\partial_t\big[\sqrt{g^{(3)}}\rho(\vec{x},t)\big]=-\partial_t[\sqrt{g^{(3)}}\,]P(\vec{x},t)\;.
\end{align} Where we assumed 
\begin{align}
T_{\mu\nu} = {\rm diag}\Big(\rho(\vec{x},t)\,,\,-g^{(3)}_{ij}(\vec{x},t)P(\vec{x},t)\Big)
\end{align}
and we used $\partial_t \,{\rm det}(g^{(3)}) = \dot g^{(3)}{}^{ij}g^{(3)}_{ij} $. If we further assume the pressure $P(\vec{x},t)$ is homogeneous, as it should be for a first-order phase transition, and integrate over comoving coordinates $\vec{x}$, we recover \eqref{eq:pdV}.

We can estimate the amount of entropy generated by slight deviations from equilibrium. The first such process is the initial rapid growth of the bubbles after their nucleation in the supercooled deconfined phase, reheating it back to temperature $T_c$ where equilibrium is restored. As a bubble of confined phase within the supercooled deconfined phase expands by a volume $\Delta V$, the entropy in that volume drops by an amount $- L\Delta V/T_c$, where $L$ is the latent heat per unit volume of the phase transition. The corresponding entropy drop per unit volume is $-L/T_c$. Since the confined phase is also lower in energy, the bubble will also release an amount of energy in the form of thermal energy, equal to $L\Delta V$. This will increase the total entropy by an amount $L\Delta V/T$, giving a total change in entropy $\Delta S =( L/T - L/T_c)\Delta V \simeq \epsilon L\Delta V/T_c$. 

A more careful calculation, assuming that the temperature rises continuously as the energy is released, gives a factor of a half
\begin{align}\Delta S  \simeq \frac{\epsilon L\Delta V}{2T_c}
\end{align}
where $\epsilon$ is the initial supercooling at the onset of bubble nucleation. Of course, the above calculation assumes that latent heat has only weak dependence on temperature away from $T_c$. For the remainder of the phase transition, where the supercooling is both much smaller and slowly varying, the calculation is much the same, and we find that if a fraction $\delta x$ of phase is converted to confined phase, an amount $\Delta S = \epsilon L \delta x V/T_c$ of entropy is produced within a comoving volume $V$. 

A formula for the entropy production that gives both these effects together is
\begin{align}
\Delta S \simeq \frac{L}{T_c}\int_0^1\!\epsilon(x)V(x) dx\;.
\end{align}
if the volume $V(x)$ is assumed to vary very little during the phase transition, it can be pulled out of the integral, giving $\Delta S = L V \bar\epsilon /T_c$, where $\bar \epsilon$ is the average supercooling during the PT. This finding is in agreement with the observation of Ref.~\cite{DeGrand:1984uq}.

Using this fact, and the fact that the entropy density of a system with zero chemical potential is given by
\begin{align}
    s = \frac{\rho + P}{T}\;,
\end{align}
we can calculate the ratio of initial and final volumes of a given comoving patch by using 
\begin{align}
TS_f & = (\rho_f + P_f)V_f = (\rho_i + P_i)V_i + T\Delta S \nonumber \\ &  \simeq (\rho_i + P_i + \bar\epsilon L)V_i
\end{align}
we get 
\begin{align}
\frac{V_f}{V_i} \simeq \frac{\rho_i + P_i + \bar\epsilon L}{\rho_f + P_f}\;.
\end{align}
In the confined phase, we can approximate that $\rho_f \simeq \rho_{\rm SM}$, since the glueball density $\rho_{\rm GB}\ll \rho_{\rm SM}$. In the deconfined phase, the density would then need to be $\rho_i = \rho_{\rm SM} + L$, where $L$ is the latent heat density between the two phases, computed from the lattice to be $L\sim 1.413\, T_c^4$~\cite{Lucini:2005vg}. The energy in latent heat is roughly a third of what would be expected from the Stefan-Boltzmann law, $\rho_g = \pi^2/30\times 2(N_c^2-1)T^4_c$, the difference being due to dynamical effects. 

It is possible that these same dynamical effects near the phase transition cause the glueball density to be slightly higher by the same amount than its non-interacting equilibrium value, and the gluon density to be slightly higher such that the difference is still $L$, but we assume these effects to be small. 

For the pressure in both phases, we use the fact that the pressure is constant throughout the phase transition. In particular, lattice computations \cite{Giusti_2017} indicate that the pressure of the strong sector at the phase transition is much less than the energy density, $P\ll L$. We are therefore justified in taking $P\simeq P_{\rm SM}=\rho_{\rm SM}/3$ both before and after the phase transition. Using this, and assuming the supercooling $\bar\epsilon$ to be small \cite{Asadi:2021pwo}, we find a simple expression for the scale factor at the end of the phase transition:
\begin{align}
    \frac{a_f}{a_i}\simeq \left(1+\frac{3L}{4\rho_{\rm SM}}\right)^{1/3}\!\!\!\!= \left(1+\frac{3\cdot1.413}{4\cdot106.75\,\pi^2/30}\right)^{1/3}\!\!\!\!\simeq 1.01\;\,.
\end{align}

Furthermore, assuming a deSitter type evolution, we find that the time scale of the PT is $t_{\rm PT} \approx 0.01/H$, i.e. of order $1\%$ of the Hubble time, which is consistent with the numerical findings in~Ref.~\cite{Asadi:2021pwo}. 

The crucial insight we gained is that the PT is a relatively slow process,  which means that processes that are not frozen out at $T = T_c$, and therefore fast at the time of the PT,  will be efficient in maintaining local thermal equilibrium throughout the PT.

\section{Dynamic Approach to the Phase Transition}
\label{sec:appxDyn}

In addition to the pure thermodynamic discussion, we provide an approximation that allows us to estimate the number density of glueballs that are produced in the PT from the dynamical point of view. 
Given the complicated strong dynamics during the PT, the arguments in this appendix should be understood to yield only a rough approximation for the value of $R$, while a detailed numerical simulation is in order for determining its exact value.

The basic picture is that glueballs are produced in $2 \rightarrow 2$ reactions from dark sector gluons, at the boundary between the two phases. 
The  process can be viewed as a two step reaction. First, at the wall between the deconfined and confined phase, two gluons are converted into a glueball. This part of the reaction is a non-perturbative process, but its rate factor scales as $1/T_c^2$. In the second step, in order to conserve momentum, a glueball splits into two glueball final states. As known from the effective Lagrangian for glueballs, the coupling strength of the three glueball interaction is large, and we only pay a price from the kinematic suppression of producing two massive particles. 

Thus this process has a reaction rate density 
\begin{align}
    \gamma_{\rm GB} = n_g^2 \langle \sigma_{\rm GB} v \rangle
\end{align}
where the reaction rate factor is given by
\begin{align}
  \langle \sigma_{\rm GB} v \rangle \approx 1/T_c^2 \exp{ \left[-2 m_{\rm GB}/T_c \right] }\,.
\end{align}

As discussed in Ref.~\cite{Asadi:2021bxp}, the bubbles nucleate and grow in size until the bubble walls collide, at a typical radius $R_1$, called the percolation radius. After that, the pockets of deconfined phase shrink, therefore, the largest relative volume that is occupied by bubble walls, of the region that contains one bubble, is around the time of percolation. Therefore, the most efficient glueball production takes place when the bubbles have a size $R \sim R_1$, and we focus here on that phase. 

The relative volume that is occupied by the walls in a region that contains one bubble at this time is thus given by 
\begin{align}
    f_V  = \frac{V_{\rm walls}}{V_{\rm bubble}} = \frac{ 4 \pi T_c^{-1} R_1^2}{\frac{4 \pi}{3}\, R_1^3} = 3 \left( \frac{10^4 \, T_c}{M_{\rm pl}} \right)^{2/3}\,.
\end{align}
here the expression for $R_1$ of Ref.~\cite{Asadi:2021bxp} was used. The total number of glueballs that are produced on average per volume in the PT is thus
\begin{align}
    n_{\rm GB} = \gamma_{\rm GB} \, f_V\,  t_{\rm prod} \,,
\end{align}
where $t_{\rm prod}$ is the time scale over which production is efficient. Since the glueball production process is dominated around the bubble percolation, the production time scale is given by the time the system spends around the percolation period $ t_{\rm prod} \approx R_1/v_w$. This gives us an estimate for the produced glueball number density of
\begin{align}
\label{eq:nGBexcess}
 n_{\rm GB} \approx T_c^3 \frac{ \exp{ \left[- 2 m_{\rm GB}/T_c \right] }}{v_w}\,,
\end{align}
where $v_w \approx 0.2 \left( T_c/M_{\rm pl}\right)^{0.2}$, as shown in Ref.~\cite{Asadi:2021bxp}.

We find that while the mass of the glueball leads to a suppressed production probability, the relatively slow advance of the wall (which produces the glueballs) partially compensates that suppression. The glueball number densities that are achieved are comparable or larger than the thermally expected abundances, used to derive Eq.~\eqref{eq:Rmax}. 

Comparing the energy density $m_{\rm GB} n_{\rm GB}$ implied by Eq.~\eqref{eq:nGBexcess} to the latent heat per volume $L$ released in the PT, one can think of the excess glueballs (with respect to their equilibrium value) as being produced by a small fraction $f\sim 10^{-2}$ of the total latent heat. In chemical equilibrium, this latent heat would be entirely released in the form of heat. In this case, however, where the rate of $3\rightarrow 2$ interactions is fast but finite, we can think of the PT happening in steps, where the excess glueballs are initially produced, and subsequently eliminated by $3\rightarrow 2$ interactions, driving the glueball number density to its equilibrium value and releasing heat. 
Comparing the rate of $3\to 2$ (at densities at or above the equililbrium value $n_{\rm GB}\simeq (m_{\rm GB}T_c/2\pi)^{3/2}\exp(-m_{\rm GB}/T_c)\,$) to the total time for the PT, $\mathcal{O}(10^{-2}-10^{-1})H^{-1}$, shows that the rate of $3\rightarrow2$ is indeed fast, and that we can therefore assume the equilibrium value for $n_{\rm GB}$.

\section{Heavier Glueballs}
\label{sec:appxGB}

Throughout this work, we only kept the lightest glueball in our treatment and claimed heavier ones do not affect our analysis in a significant way. In this appendix we justify this claim. 

In Ref.~\cite{Forestell:2016qhc} it is argued that immediately after the PT, if enough glueballs are produced so that various number changing processes are in equilibrium, the abundance of heavier CP-even glueballs deplete very rapidly. 
To see this, we should keep in mind that the heavier CP-even glueballs coannihilate via $2\rightarrow 2$ processes to the $0^{++}$ state, while the $0^{++}$ abundance goes through freezeout via $3\rightarrow 2$ processes with longer timescales.

The CP-odd glueballs also have a far lower abundance compared to the $0^{++}$ glueball after the PT \cite{Forestell:2016qhc}. 
As a result, before their decay, the lightest glueballs ($0^{++}$) dominate the dynamics of the glueball bath such that we can study a simplified model with only these glueballs in the spectrum.

Neglecting the $Z'$ mass, we can use the results of Ref.~\cite{Juknevich:2009gg} to show that in our model 
\begin{eqnarray}
    \label{eq:GBsdecayratios}
    \tau_{0^{++}} \approx 1.766 \tau_{0^{-+}} \approx 0.040 \tau_{2^{++}} \approx 0.004 \tau_{2^{-+}},
\end{eqnarray}
while other glueballs decay to one of these glueballs with a far shorter lifetimes when $\alpha_{Z'}$ is perturbative. Combined with the above argument, we find that we can only keep $0^{++}$ and $2^{\pm +}$ (that are more long-lieved than $0^{++}$) throughout our study.

When glueballs decay to SM, the entropy they inject into SM can be calculated as 
\begin{equation}
    S = \int \frac{dQ}{T},
    \label{eq:dSeq}
\end{equation}
where $dQ$ is the heat they inject into SM during an infinitesimal time interval and $T$ is the temperature at which this injection takes place. 
Since the $0^{++}$ states have a much higher energy density than $2^{\pm +}$ ones, the only way the latter can have significant entropy injection into SM is that the injection takes place at lower temperatures. 
Equation~\eqref{eq:GBsdecayratios} suggests that indeed $2^{\pm +}$ entropy injection happens at $T_{\tau_{2^{\pm +}}} \sim 10^{-1} T_{\tau_{GB}}$; nonetheless, since the total energy $Q$ stored in $2^{\pm +}$ states is much lower than the $0^{++}$ states~\cite{Forestell:2016qhc}, $T_{\tau_{2^{\pm +}}} \sim 10^{-1} T_{\tau_{GB}}$ is not low enough to change the total injected entropy into SM perceptibly. 
Hence, keeping only the $0^{++}$ state is a good approximation for calculating the injected entropy into the SM bath.

The only relevant effect of the heavier glueballs is that if they decay too late, they can be in conflict with stringent bounds from BBN \cite{Jedamzik:2006xz,Kawasaki:2017bqm}. These bounds are avoided by demanding that all the glueballs decay before the onset of BBN, i.e. $t \sim 1$s. Since the longest-living glueball is $2^{-+}$, this translates into $\tau_{2^{-+}} \lesssim 1$s; when combined with Eq.~\eqref{eq:GBsdecayratios}, this suggests 
\begin{equation}
\tau_{0^{++}} \lesssim 10^{-2} ~\mathrm{s}.
\label{eq:tauGBappx}
\end{equation}
We use this upper bound on the lightest glueball lifetime in our analysis. 
A more careful treatment can allow for $\mathcal{O}(1)$ change in this bound, which in turn merely gives rise to $\mathcal{O}(1)$ change in the dilution factor from the entropy injection. This is sub-dominant to various sources of uncertainty in our calculation of DM relic abundance during the PT \cite{Asadi:2021pwo} and is neglected in this work.


\bibliographystyle{utphys_modified}
\bibliography{bib}

\end{document}